# Amplitude-Aided 1-Bit Compressive Sensing Over Noisy Wireless Sensor Networks

Ching-Hsien Chen and Jwo-Yuh Wu*

***Abstract*-One-bit compressive sensing (CS) is known to be particularly suited for resource-constrained wireless sensor networks (WSNs). In this paper, we consider 1-bit CS over noisy WSNs subject to channel-induced bit flipping errors, and propose an *amplitude-aided* signal reconstruction scheme, by which (i) the representation points of local binary quantizers are designed to minimize the loss of data fidelity caused by local sensing noise, quantization, and bit sign flipping, and (ii) the fusion center adopts the conventional $\ell_1$ – minimization method for sparse signal recovery using the decoded and de-mapped binary data. The representation points of binary quantizers are designed by minimizing the mean square error (MSE) of the net data mismatch, taking into account the distributions of the nonzero signal entries, local sensing noise, quantization error, and bit flipping; a simple closed-form solution is then obtained. Numerical simulations show that our method improves the estimation accuracy when SNR is low or the number of sensors is small, as compared to state-of-the-art 1-bit CS algorithms relying solely on the sign message for signal recovery.**

***Index Terms:* compressive sensing; quantization; wireless sensor networks.**

## I. INTRODUCTION

Compressive sensing (CS) provides a new paradigm for sub-Nyquist signal processing capable of reducing the data acquisition and storage costs [1-2]. While many existing studies of CS assumed real-valued measurements, the needs of high-speed digital transmission and processing have stimulated the development of CS schemes using quantized measurements [3], in certain cases even with one-bit coarse quantization to ease hardware implementation, aka., the 1-bit CS [4-7]. Recently, integration of CS into the design of wireless sensor networks (WSNs) has received considerable attention [8-9]. The 1-bit CS is particularly attractive in this scenario, due to its capability of reducing the communication and computational costs of local sensors [10-11]. In practical WSNs, communication link errors caused by channel fading and noise are unavoidable. Hence, transmission of the one-bit message from each local node to the fusion center (FC) will be subject to random bit flipping. The design of sparse signal reconstruction algorithms for 1-bit CS by further taking into account the effect of bit flipping has been recently addressed in, e.g., [5-7].

In this paper we study 1-bit CS over a noisy WSN, which employs the following data processing protocol: (i) each sensor observes a common $K$-sparse signal vector corrupted with noise, (ii) compresses its observation into a scalar, (iii) quantizes the compressed measurement into one bit, and (iv) transmits this binary information in parallel through a noisy communication link, modeled as a binary symmetric

This work is sponsored by the National Science Council of Taiwan under grants NSC 102-2221-E-009-019-MY3 and MOST 103-2221-E-009-050-MY3, by the Ministry of Education of Taiwan under the MoE ATU Program.

The authors are with the Department of Electrical and Computer Engineering, National Chiao Tung University, Taiwan; tel: 886-35-712121, ext. 54524; fax: 886-35-710116. Emails: tshs0821@hotmail.com; jywu@cc.nctu.edu.tw.

* Contact author.

channel (BSC), to the FC for global signal reconstruction. In contrast to all existing 1-bit CS methods using exclusively the sign information for signal reconstruction, we propose an *amplitude-added* sparse signal recovery scheme to combat the bit flipping caused by BSC. In our approach, the representation points of the local binary quantizers are designed by minimizing the mean square error (MSE) of the data mismatch at the FC upon bit decoding and magnitude de-mapping. Then, using the de-mapped binary data, conventional signal reconstruction algorithms based on real-valued measurements, such as the standard $\ell_1$-minimization method [12], is employed at the FC for signal reconstruction. The adopted MSE metric takes into account the distributions of nonzero entries of the sparse signal, local sensing noise, quantization error, and the bit flipping effect. An analytic formula for the MSE is first derived. The optimal quantizer representation level is then obtained in closed-form, without resorting to intensive numerical optimization. Compared to current state-of-the-art schemes [5-7], the distinctive features of the proposed method can be summarized as follows.

1. Existing 1-bit CS algorithms relied fully on the received sign patterns for signal reconstruction, ignoring the average signal amplitude information revealed by the representation level. By contrast, the proposed approach with MSE-optimal binary quantizer design can leverage the signal amplitude knowledge for performance enhancement, especially in the presence of severe bit flipping caused by large data transmission errors and measurement noise.
2. Computer simulations show that the proposed method improves the reconstruction performance when the SNR is low or the total number of sensor nodes is small.

## II. SYSTEM MODEL AND BASIC ASSUMPTIONS

We consider a WSN, in which $M$ sensor nodes cooperate with a FC for estimating a common $K$-sparse vector $\mathbf{s} \in \mathbb{R}^N$, with support $\mathcal{T} \subset \{1,\cdots,N\}$ such that $|\mathcal{T}| = K$ and $M \ll N$. The received signal at the $i$th sensor node is

$$\mathbf{x}_i = \mathbf{s} + \mathbf{v}_i,\ 1 \le i \le M, \tag{2.1}$$

where $\mathbf{v}_i \sim \mathcal{N}(\mathbf{0}_N, \sigma_v^2 \mathbf{I}_N)$ is white Gaussian measurement noise. To conserve the energy and bandwidth resources, the raw sensor data $\mathbf{x}_i$ is not directly forwarded to the FC. Instead, each sensor first compresses its measurement $\mathbf{x}_i$ into a scalar as

$$z_i = \mathbf{\Phi}_i \mathbf{x}_i = \mathbf{\Phi}_i(\mathbf{s} + \mathbf{v}_i),\ 1 \le i \le M, \tag{2.2}$$

in which $\mathbf{\Phi}_i = [\mathbf{\Phi}_{i1} \cdots \mathbf{\Phi}_{iN}] \in \mathbb{R}^{1\times N}$ is the unit-norm data compression vector with $\mathbf{\Phi}_{ij} \in \mathbb{R}$ as the $j$-th entry. Afterwards, the compressed measurement $z_i$ is quantized into a binary message

$$q_i = \mathcal{Q}_i(z_i) = \begin{cases} \alpha_i, & \text{if } z_i \ge \tau_i; \\ \beta_i, & \text{if } z_i < \tau_i, \end{cases} \tag{2.3}$$

according to a binary quantization rule $Q_i(\cdot)$, with $\tau_i$ as the threshold and $\{\alpha_i, \beta_i\}$ as the representation points. Each of the quantized messages $q_i$, $1 \le i \le M$, is encoded into one bit, and is then transmitted over $M$ parallel channels to the FC. To account for the effects of fading and channel noise, each communication link is modeled as a BSC with cross-over probability $P_{e,i}$, $1 \le i \le M$. Let $\hat{q}_i \in \{\alpha_i, \beta_i\}$ be the quantized data received at the FC (upon bit decoding and de-mapping) from the $i$th sensor node, $1 \le i \le M$; due to imperfect decoding, the event that $\hat{q}_i \neq q_i$ occurs with a probability $P_{e,i}$. Collecting $\hat{q}_i$'s received at the FC into a vector, we write

$$\mathbf{y} = [\hat{q}_1 \cdots \hat{q}_M]^T = \underbrace{\begin{bmatrix} \mathbf{\Phi}_1^T & \cdots & \mathbf{\Phi}_M^T \end{bmatrix}^T}_{\triangleq \mathbf{\Phi}} \mathbf{s} + \mathbf{w}, \tag{2.4}$$

where $\mathbf{\Phi} \in \mathbb{R}^{M \times N}$ is the system sensing matrix, and $\mathbf{w} \in \mathbb{R}^M$ is the aggregate noise vector accounting for the local sensing noise, quantization error, and bit flipping. In this paper, we propose an amplitude-aided sparse signal reconstruction scheme using the de-mapped binary data $\mathbf{y}$ given in (2.4). The following assumptions are made in the sequel.

***Assumption 1:*** The $K$ nonzero entries of the sparse signal $\mathbf{s}$ are i.i.d. Gaussian random variables, i.e., $s_k \sim \mathcal{N}(0, \sigma_s^2)$ for all $k \in \mathcal{T}$, and are independent with the local sensing noise $\mathbf{v}_i$'s. □

***Assumption 2:*** The sensing matrix $\mathbf{\Phi}$ is binary with entries $\mathbf{\Phi}_{ij} \in \{\pm 1/\sqrt{N}\}$. □

# III. AMPLITUDE-AIDED ONE-BIT CS

## *A. Proposed Approach*

The idea behind the proposed scheme is to exploit the side information about the signal magnitude in addition to the 1-bit sign message. In the context of 1-bit CS with binary quantization employed at each sensor node, a natural approach is thus to judiciously design the binary quantizer $\{\tau_i, \alpha_i, \beta_i\}$ so that the loss of data fidelity (caused by local sensing noise, signal quantization, and bit flipping) at the FC is kept to the minimum. Specifically, the proposed approach consists of the following two steps:

- A design of the local binary quantizer parameters $\{\tau_i, \alpha_i, \beta_i\}_{1 \le i \le M}$ by minimizing the MSE of the data mismatch $E\{\|\mathbf{w}\|^2\}$, where the expectation is taken with respect to the distributions of the nonzero signal entries, local sensing noise, quantization error, and bit flipping.
- With the optimal quantizers deployed at local nodes, the FC performs the $\ell_1$-minimization algorithm [12] for sparse signal recovery[1].

1. To guarantee stable signal reconstruction using the $\ell_1$-minimization algorithm [12], a commonly-used sufficient condition is that the sensing matrix $\mathbf{\Phi}$ satisfies the restricted isometry property (RIP) of order $2K$ [12]. Notably, the RIP condition holds with an overwhelmingly large probability for binary sensing matrices with entries drawn from i.i.d. symmetric Bernoulli distribution [2].

A key step of the proposed approach is to design the MSE-optimal binary quantizer, as to be discussed in the next two subsections.

*B. MSE-Optimal Quantizer: Design Formulation*

At the $i$th node, the input to the quantizer is $z_i = \mathbf{\Phi}_i(\mathbf{s} + \mathbf{v}_i)$, which is Gaussian random variable with mean $E\{z_i\} = \mathbf{\Phi}_i(E\{\mathbf{s}\} + E\{\mathbf{v}_i\}) = 0$ (recall both the signal $\mathbf{s}$ and noise $\mathbf{v}_i$ are zero mean). The variance of $z_i$ is

$$\begin{aligned} E\{|z_i|^2\} &= \mathbf{\Phi}_i E\{(\mathbf{s}+\mathbf{v}_i)(\mathbf{s}+\mathbf{v}_i)^T\}\mathbf{\Phi}_i^T \overset{(a)}{=} \mathbf{\Phi}_i E\{\mathbf{s}\mathbf{s}^T\}\mathbf{\Phi}_i^T + \mathbf{\Phi}_i E\{\mathbf{v}_i\mathbf{v}_i^T\}\mathbf{\Phi}_i^T \\ &\overset{(b)}{=} \sigma_s^2 \sum_{j\in\mathcal{T}} |\mathbf{\Phi}_{ij}|^2 + \mathbf{\Phi}_i E\{\mathbf{v}_i\mathbf{v}_i^T\}\mathbf{\Phi}_i^T \overset{(c)}{=} \underbrace{\frac{K\sigma_s^2}{N}}_{\triangleq\sigma^2} + \mathbf{\Phi}_i E\{\mathbf{v}_i\mathbf{v}_i^T\}\mathbf{\Phi}_i^T \overset{(d)}{=} \sigma^2 + \sigma_v^2, \end{aligned} \tag{3.1}$$

where (a) follows since $\mathbf{s}$ and $\mathbf{v}_i$ are independent, (b) follows from Assumption 1, (c) is true since $\mathbf{\Phi}_{ij} \in \{\pm 1/\sqrt{N}\}$ and $|\mathcal{T}| = K$, and (d) holds by using $E\{\mathbf{v}_i\mathbf{v}_i^T\} = \sigma_v^2\mathbf{I}$ and $\|\mathbf{\Phi}_i\|^2 = 1$. With (3.1), it can be concluded that $z_i \sim \mathcal{N}(0, \sigma^2 + \sigma_v^2)$. Due to symmetry of the Gaussian density, we accordingly have $\tau_i = 0$ and $\beta_i = -\alpha_i$, thereby the quantization rule (2.3) simplified to

$$q_i = \mathcal{Q}_i(z_i) = \begin{cases} \alpha_i, & \text{if } z_i \geq 0; \\ -\alpha_i, & \text{if } z_i < 0. \end{cases} \tag{3.2}$$

Henceforth, the design of the binary quantizer employed at the $i$th node amounts to the design of a single representation point $\alpha_i$; without loss of generality, $\alpha_i > 0$ is assumed in the sequel. Hence, our purpose is to find

$$(\bar{\alpha}_1, \cdots, \bar{\alpha}_M) = \underset{\alpha_i > 0,\ 1 \leq i \leq M}{\arg\min}\ E\{\|\mathbf{w}\|^2\}, \tag{3.3}$$

where the expectation takes into account the distributions of the signal entries, local sensing noise, quantization error, and bit flipping.

*C. MSE-Optimal Quantizer: Solution*

To solve problem (3.3), we first derive an analytic formula for $E\{\|\mathbf{w}\|^2\}$. Let us write $\mathbf{w} = [w_1 \cdots w_M]^T$, where $w_i \in \mathbb{R}$, and then

$$\begin{aligned} E\{\|\mathbf{w}\|^2\} &= \sum_{i=1}^{M} E\{|w_i|^2\} \overset{(a)}{=} \sum_{i=1}^{M} E\{|\hat{q}_i - \mathbf{\Phi}_i\mathbf{s}|^2\} \\ &= \sum_{i=1}^{M} \Big[ E\{|\hat{q}_i - \mathbf{\Phi}_i\mathbf{s}|^2 \mid \mathbf{\Phi}_i\mathbf{s} \geq 0\} \Pr\{\mathbf{\Phi}_i\mathbf{s} \geq 0\} + E\{|\hat{q}_i - \mathbf{\Phi}_i\mathbf{s}|^2 \mid \mathbf{\Phi}_i\mathbf{s} < 0\} \Pr\{\mathbf{\Phi}_i\mathbf{s} < 0\} \Big], \end{aligned} \tag{3.4}$$

where (a) follows from (2.4). Notably, conditioned on $\mathbf{\Phi}_i\mathbf{s} \geq 0$ and since $\hat{q}_i \in \{\pm\alpha_i\}$, it can be directly deduced that $|\hat{q}_i - \mathbf{\Phi}_i\mathbf{s}|^2 = |\alpha_i - \mathbf{\Phi}_i\mathbf{s}|^2$ if $\mathrm{sgn}(\mathbf{\Phi}_i\mathbf{s}) = \mathrm{sgn}(\hat{q}_i)$, i.e., $\mathbf{\Phi}_i\mathbf{s}$ and $\hat{q}_i$ have the same sign, and $|\hat{q}_i - \mathbf{\Phi}_i\mathbf{s}|^2 = |\alpha_i + \mathbf{\Phi}_i\mathbf{s}|^2$ if $\mathrm{sgn}(\mathbf{\Phi}_i\mathbf{s}) \neq \mathrm{sgn}(\hat{q}_i)$; conditioned on $\mathbf{\Phi}_i\mathbf{s} < 0$, we instead have

$|\hat{q}_i - \mathbf{\Phi}_i\mathbf{s}|^2 = |\alpha_i + \mathbf{\Phi}_i\mathbf{s}|^2$ if $\mathrm{sgn}(\mathbf{\Phi}_i\mathbf{s}) = \mathrm{sgn}(\hat{q}_i)$, and $|\hat{q}_i - \mathbf{\Phi}_i\mathbf{s}|^2 = |\alpha_i - \mathbf{\Phi}_i\mathbf{s}|^2$ if $\mathrm{sgn}(\mathbf{\Phi}_i\mathbf{s}) \neq \mathrm{sgn}(\hat{q}_i)$ Hence, if we write

$$P_i^+ \triangleq \Pr\{\mathrm{sgn}(\mathbf{\Phi}_i\mathbf{s}) \neq \mathrm{sgn}(\hat{q}_i) \mid \mathbf{\Phi}_i\mathbf{s} \geq 0\} \text{ and } P_i^- \triangleq \Pr\{\mathrm{sgn}(\mathbf{\Phi}_i\mathbf{s}) \neq \mathrm{sgn}(\hat{q}_i) \mid \mathbf{\Phi}_i\mathbf{s} < 0\}, \tag{3.5}$$

it then follows

$$E\{|\hat{q}_i - \mathbf{\Phi}_i\mathbf{s}|^2 \mid \mathbf{\Phi}_i\mathbf{s} \geq 0\} = |\mathbf{\Phi}_i\mathbf{s} - \alpha_i|^2 \times (1 - P_i^+) + |\mathbf{\Phi}_i\mathbf{s} + \alpha_i|^2 \times P_i^+ \tag{3.6}$$

and

$$E\{|\hat{q}_i - \mathbf{\Phi}_i\mathbf{s}|^2 \mid \mathbf{\Phi}_i\mathbf{s} < 0\} = |\mathbf{\Phi}_i\mathbf{s} + \alpha_i|^2 \times (1 - P_i^-) + |\mathbf{\Phi}_i\mathbf{s} - \alpha_i|^2 \times P_i^-. \tag{3.7}$$

From (3.4)~(3.7), a key step to find the formula of $E\{\|\mathbf{w}\|^2\}$ is thus to determine the conditional probabilities $P_i^+$ and $P_i^-$. This is done in the next lemma.

***Lemma 3.1:*** The following result holds:

$$P_i^+ = P_i^- = P_{e,i} + (1 - 2P_{e,i}) \times Q\left(|\mathbf{\Phi}_i\mathbf{s}| / \sigma_v\right), \tag{3.8}$$

where $P_{e,i}$ is the cross-over probability of the $i$th BSC and $Q(\cdot)$ is the standard $Q$-function [13]. □

*[Proof]:* See Appendix A. □

With the aid of Lemma 3.1, a closed-from formula for $E\{\|\mathbf{w}\|^2\}$ is given in the following theorem.

***Theorem 3.2:*** The mean square error of the mismatched term $\mathbf{w}$ can be expressed as

$$E\{\|\mathbf{w}\|^2\} = \sum_{i=i}^{M}\left[\left(\alpha_i - \sqrt{\frac{2\sigma^2(1-2P_{e,i})^2}{\pi(1+(\sigma_v/\sigma)^2)}}\right)^2 + \left(\sigma^2 - \frac{2\sigma^2(1-2P_{e,i})^2}{\pi(1+(\sigma_v/\sigma)^2)}\right)\right], \tag{3.9}$$

where $\sigma^2$ is defined in (3.1). □

*[Proof]:* See Appendix B. □

With (3.9), minimization of $E\{\|\mathbf{w}\|^2\}$ over $\alpha_i$'s can thus be done on a node-by-node basis, and the optimal solutions are immediately obtained as

$$\bar{\alpha}_i = \sqrt{\frac{2\sigma^2(1-2P_{e,i})^2}{\pi \cdot (1+(\sigma_v/\sigma)^2)}}, \quad 1 \leq i \leq M. \tag{3.10}$$

The resultant minimal MSE is

$$\overline{MSE} = \sum_{i=1}^{M}\left(\sigma^2 - \frac{2\sigma^2(1-2P_{e,i})^2}{\pi(1+(\sigma_v/\sigma)^2)}\right). \tag{3.11}$$

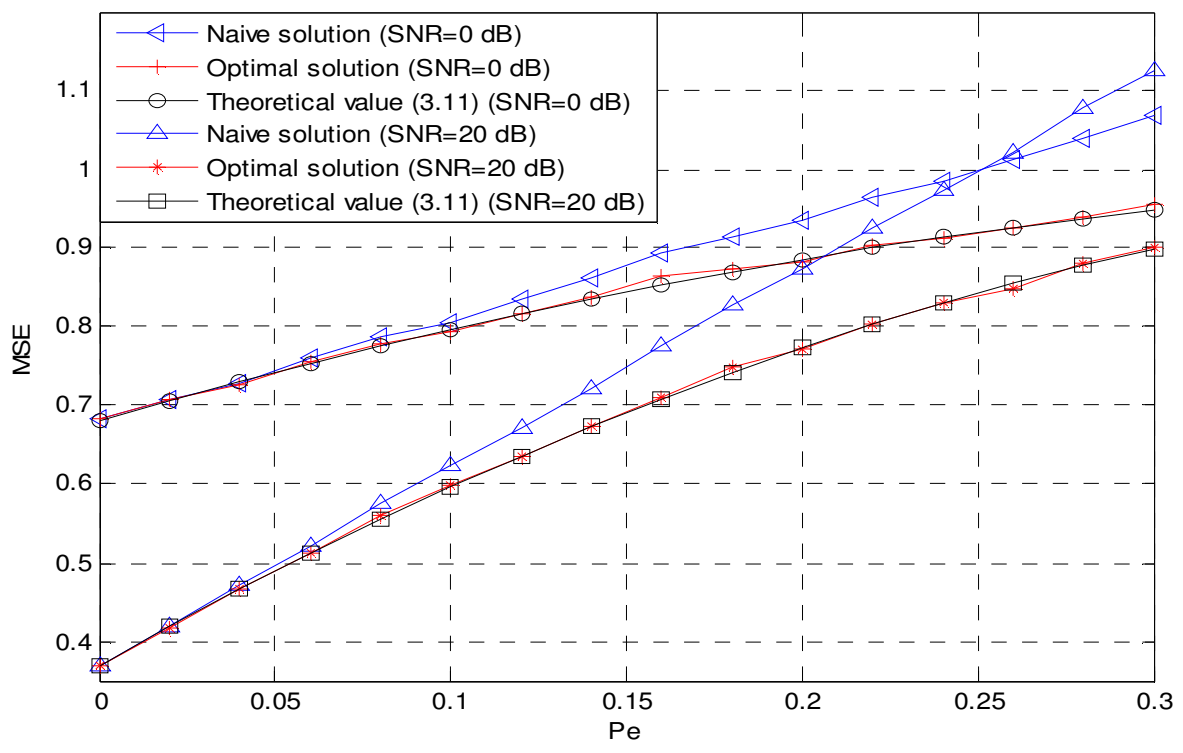

Fig. 1. MSE of the proposed optimal representation level (3.10) and the naive solution.

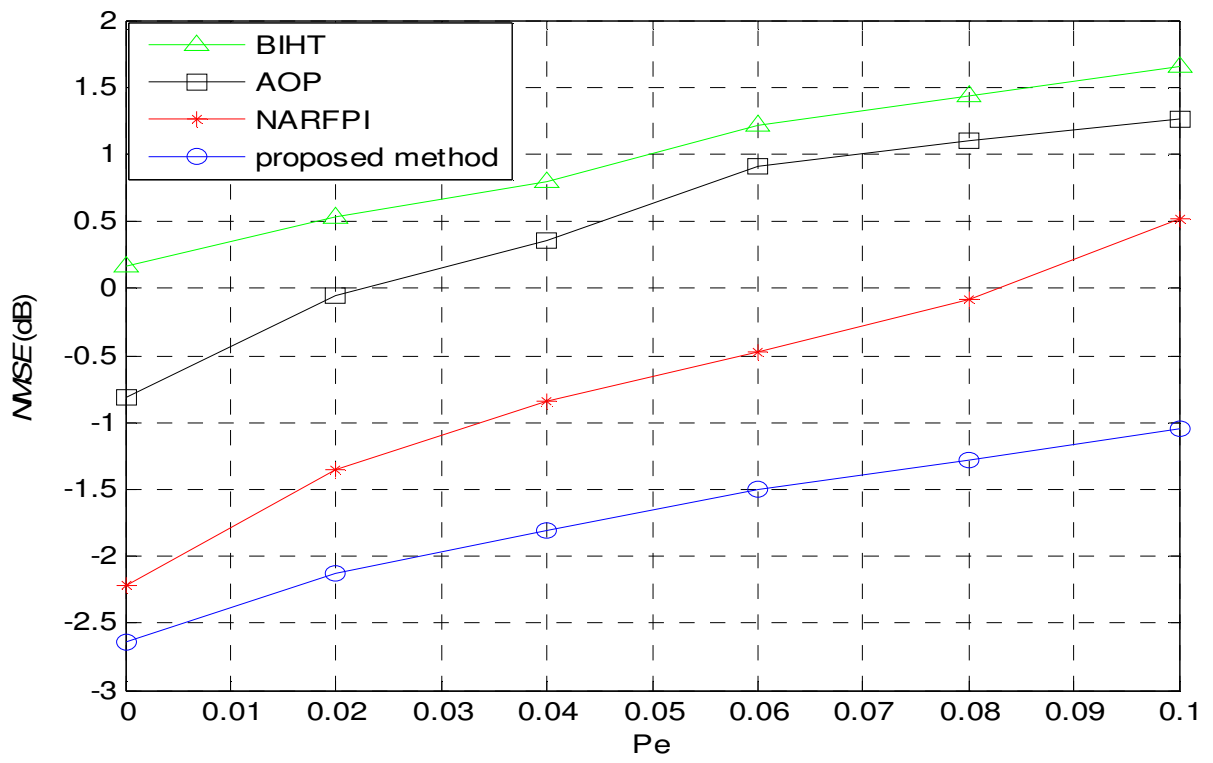

Fig. 2. NMSE of four signal reconstruction methods as a function of the bit flipping probability (SNR=10 dB).

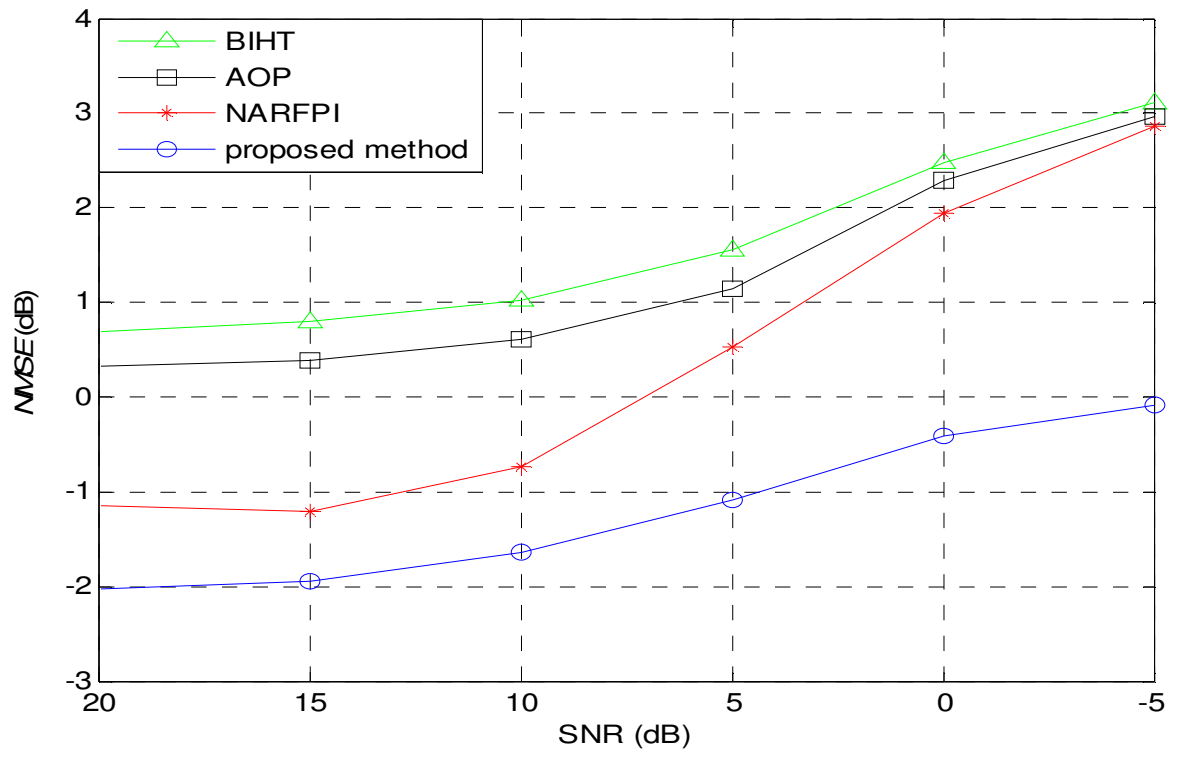

Fig. 3. NMSE of four signal reconstruction methods as a function of SNR ( $P_e$=0.05 ).

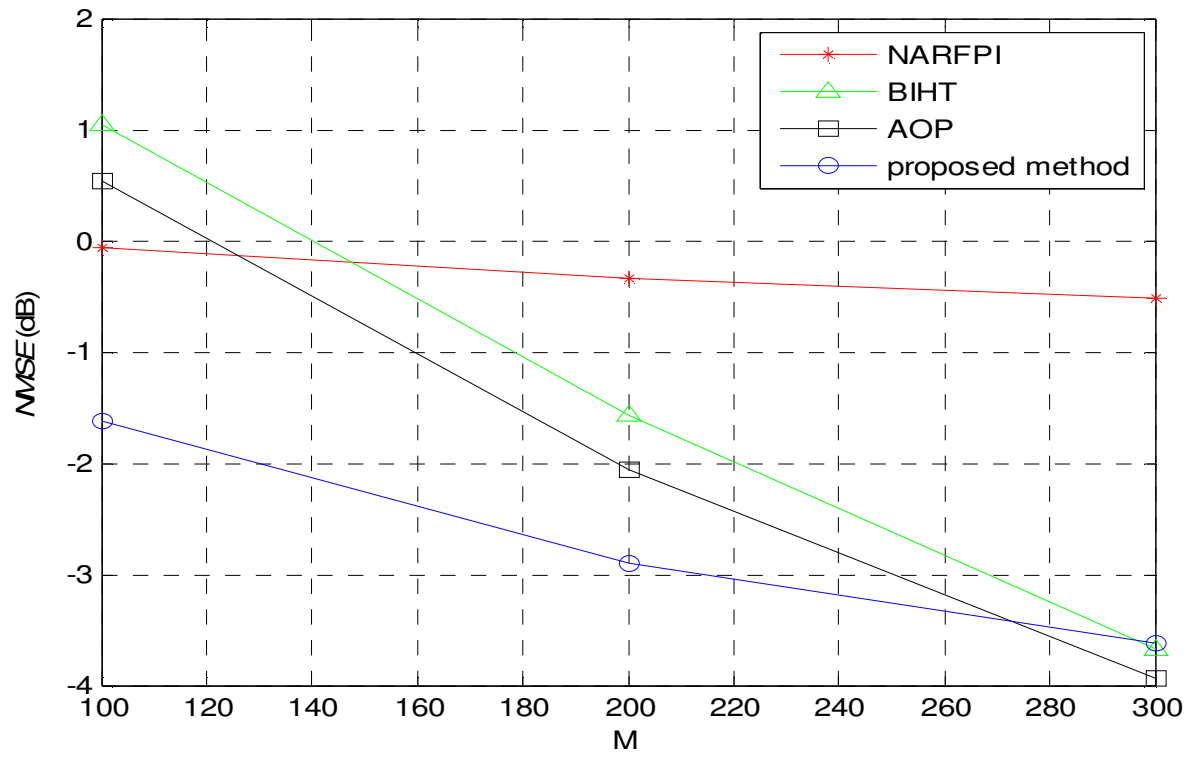

Fig. 4. NMSE of four signal reconstruction methods as a function of the total number of sensors (SNR=10 dB and $P_e$=0.05 ).

# IV. RESULTS AND DISCUSSIONS

We use numerical simulations to illustrate the performance of the proposed method. For simplicity of illustration, we assume the homogeneous link condition, i.e., $P_{e,i} = P_e$, for all $1 \le i \le M$; the SNR of the local measurement is defined as $SNR \triangleq E\{\|\mathbf{s}\|^2\} / E\{\|\mathbf{v}_i\|^2\}$. With $M = 100$, $N = 1000$, and $K = 10$, Figure 1 first compares the proposed optimal representation level (3.10) with the naive solution obtained without considering the bit flipping effect, that is, the one obtained by setting $P_e = 0$ in (3.10), leading to $\tilde{\alpha}_i = \sqrt{\dfrac{2K\sigma_s^2}{\pi N(1+(N\sigma_v / K\sigma_s)^2)}}$. For both solutions, the empirical MSE corresponding to, respectively, $SNR = 0$ dB and 20 dB are plotted as a function of the cross-over probability $P_e$; for the proposed solution (3.10), the theoretical values of MSE in (3.11) are also included in the figure. We can see from the figure that (i) the proposed method outperforms the naive solution, and (ii) our analytic results (3.11) are in close agreement with the simulated outcome. To further illustrate the signal reconstruction performance, the following four sparse signal reconstruction algorithms are considered, namely, the conventional $\ell_1$ −minimization combined with the proposed optimal quantizer (3.10), the binary iterative hard thresholding (BIHT) algorithm [5], adaptive outlier pursuit (AOP) [6], and noise-adaptive renormalized fixed point iteration (NARFPI) [7]. The quality of signal recovery is assessed by the *normalized mean square error*, defined to be $NMSE \triangleq E\{\|\mathbf{s}-\hat{\mathbf{s}}\|^2\}/E\{\|\mathbf{s}\|^2\}$, where $\hat{\mathbf{s}}$ is the reconstructed sparse signal at the FC. For SNR=10 dB, Figure 2 compares the achieved $NMSE$ of all methods as a function of the cross-over probability $P_e$, showing that the proposed method is quite robust against the sign flipping. With $P_e = 0.05$, Figure 3 then plots the $NMSE$ as a function of SNR; as we can see, the proposed method improves the reconstruction performance in the low SNR regime. For $P_e = 0.05$ and SNR=10 dB, Figure 4 plots $NMSE$ as a function of *M*, the total number of sensors; our method is seen to yield the best performance when $M$ is small. The above results confirm that the proposed amplitude-assisted scheme is attractive in a harsh sensing environment (with low SNR) or in a small-size network with reduced implementation cost.

# APPENDIX

*A. Proof of Lemma 3.1*

Conditioned on fixed $\mathbf{\Phi}_i\mathbf{s}$, we have $\mathbf{\Phi}_i\mathbf{s}+\mathbf{\Phi}_i\mathbf{v}_i \sim \mathcal{N}(\mathbf{\Phi}_i\mathbf{s},\sigma_v^2)$, $1 \le i \le M$. We first focus on the case $\mathbf{\Phi}_i\mathbf{s} \ge 0$, and consider the two events

$$\mathcal{A}_1 \triangleq \{\mathbf{\Phi}_i\mathbf{s}+\mathbf{\Phi}_i\mathbf{v}_i \ge 0 \mid \mathbf{\Phi}_i\mathbf{s} \ge 0\} \quad \text{and} \quad \mathcal{A}_2 \triangleq \{\mathbf{\Phi}_i\mathbf{s}+\mathbf{\Phi}_i\mathbf{v}_i < 0 \mid \mathbf{\Phi}_i\mathbf{s} \ge 0\}, \tag{A.1}$$

with probabilities $\Pr\{\mathcal{A}_1\} = Q(-\mathbf{\Phi}_i\mathbf{s}/\sigma_v)$ and $\Pr\{\mathcal{A}_2\} = Q(\mathbf{\Phi}_i\mathbf{s}/\sigma_v)$. Then

$$\begin{aligned}
P_i^+ &= \Pr\{\mathrm{sgn}(\mathbf{\Phi}_i\mathbf{s}_i) \ne \mathrm{sgn}(\hat{q}_i) \mid \mathbf{\Phi}_i\mathbf{s} \ge 0\} \\
&= \Pr\{\mathcal{A}_1 \cap \{\hat{q}_i \ne q_i\} \mid \mathbf{\Phi}_i\mathbf{s} \ge 0\} + \Pr\{\mathcal{A}_2 \cap \{\hat{q}_i = q_i\} \mid \mathbf{\Phi}_i\mathbf{s} \ge 0\} \\
&\overset{(a)}{=} \Pr\{\mathcal{A}_1 \mid \mathbf{\Phi}_i\mathbf{s} \ge 0\} \times \Pr\{\{\hat{q}_i \ne q_i\} \mid \mathbf{\Phi}_i\mathbf{s} \ge 0\} + \Pr\{\mathcal{A}_2 \mid \mathbf{\Phi}_i\mathbf{s} \ge 0\} \times \Pr\{\{\hat{q}_i = q_i\} \mid \mathbf{\Phi}_i\mathbf{s} \ge 0\} \\
&\overset{(b)}{=} \Pr\{\mathcal{A}_1 \mid \mathbf{\Phi}_i\mathbf{s} \ge 0\} \times \Pr\{\hat{q}_i \ne q_i\} + \Pr\{\mathcal{A}_2 \mid \mathbf{\Phi}_i\mathbf{s} \ge 0\} \times \Pr\{\hat{q}_i = q_i\} \\
&= Q(-\mathbf{\Phi}_i\mathbf{s}/\sigma_v) \times P_{e,i} + Q(\mathbf{\Phi}_i\mathbf{s}/\sigma_v) \times (1-P_{e,i}) \\
&= [1 - Q(\mathbf{\Phi}_i\mathbf{s}/\sigma_v)] \times P_{e,i} + Q(\mathbf{\Phi}_i\mathbf{s}/\sigma_v) \times (1-P_{e,i}) \\
&= P_{e,i} + (1-2P_{e,i})Q(\mathbf{\Phi}_i\mathbf{s}/\sigma_v) = P_{e,i} + (1-2P_{e,i})Q(|\mathbf{\Phi}_i\mathbf{s}|/\sigma_v),
\end{aligned} \tag{A.2}$$

where (a) and (b) hold since the channel flipping effect is independent of the local sensing noise. $P_i^-$ can be obtained in a similar way owing to the symmetric nature of the Gaussian density, and therefore the first equality in (3.8) holds. The proof is completed. □

*B. Proof of Theorem 3.2*

We need the following lemma to prove (3.9).

***Lemma A.1:*** $\int_0^\infty tQ(t/\sigma_v) \times \frac{2\exp(-t^2/(2\sigma^2))}{\sqrt{2\pi\sigma^2}}dt = \sqrt{\sigma^2/(2\pi)}\left(1-\sqrt{\sigma^2/(\sigma^2+\sigma_v^2)}\right)$.

*[Proof]:*

$$\int_0^\infty tQ\left(t/\sigma_v\right)\times\frac{2\exp(-t^2/(2\sigma^2))}{\sqrt{2\pi\sigma^2}}dt=\int_0^\infty Q\left(t/\sigma_v\right)(-\sigma^2)d\left[\frac{2\exp(-t^2/(2\sigma^2))}{\sqrt{2\pi\sigma^2}}\right]$$

$$\overset{(a)}{=}\left[Q\left(t/\sigma_v\right)\cdot(-\sigma^2)\frac{2\exp(-t^2/(2\sigma^2))}{\sqrt{2\pi\sigma^2}}\right]_{t=0}^{\infty}-\int_0^\infty\frac{2\exp(-t^2/(2\sigma^2))}{\sqrt{2\pi\sigma^2}}d\left[Q(t/\sigma_v)(-\sigma^2)\right]$$

$$\overset{(b)}{=}\left[Q(t/\sigma_v)\cdot(-\sigma^2)\cdot\frac{2\exp(-t^2/(2\sigma^2))}{\sqrt{2\pi\sigma^2}}\right]_{t=0}^{\infty}-\int_0^\infty\frac{2\exp(-t^2/(2\sigma^2))}{\sqrt{2\pi\sigma^2}}\times\frac{(\sigma^2)\cdot\exp(-t^2/(2\sigma_v^2))}{\sqrt{2\pi\sigma_v^2}}dt$$

$$=(\sigma^2)\cdot\frac{1}{\sqrt{2\pi\sigma^2}}-(\sigma^2)\int_0^\infty\frac{1}{\sqrt{2\pi\sigma_v^2}}\cdot\frac{2}{\sqrt{2\pi\sigma^2}}\cdot\exp\left(-(\sigma_v^{-2}+\sigma^{-2})t^2/2\right)dt$$

$$=\sqrt{\sigma^2/(2\pi)}-\sigma^2\cdot\frac{1}{\sqrt{2\pi\sigma_v^2}}\cdot\frac{2}{\sqrt{2\pi\sigma^2}}\cdot\frac{\sqrt{2\pi}}{\sqrt{(\sigma_v^{-2}+\sigma^{-2})}}\cdot\int_0^\infty\frac{\sqrt{(\sigma_v^{-2}+\sigma^{-2})}}{\sqrt{2\pi}}\cdot\exp\left(-(\sigma_v^{-2}+\sigma^{-2})t^2/2\right)dt$$

$$=\sqrt{\sigma^2/(2\pi)}-\sqrt{\sigma^2/(2\pi\sigma_v^2\cdot(\sigma_v^{-2}+\sigma^{-2}))}$$

$$=\sqrt{\sigma^2/(2\pi)}-\sqrt{\sigma^2/(2\pi)}\cdot\sqrt{\sigma^2/(\sigma^2+\sigma_v^2)}=\sqrt{\sigma^2/(2\pi)}\cdot(1-\sqrt{\sigma^2/(\sigma^2+\sigma_v^2)}),$$

where (a) is obtained by using integration by part and (b) holds due to the fact that $\frac{d}{dt}Q(t)=\frac{-1}{\sqrt{2\pi}}e^{(-t^2/2)}$ [13]. □

*[Proof of Theorem 3.2]:* Using (3.4)~(3.7), we have

$$\begin{aligned}E\{|w_i|^2\}&=E\{|\hat{q}_i-\mathbf{\Phi}_i\mathbf{s}|^2\}\\&=E\{|\mathbf{\Phi}_i\mathbf{s}-\alpha_i|^2\cdot(1-P_i^+)+|\mathbf{\Phi}_i\mathbf{s}+\alpha_i|^2\cdot P_i^+\big|\mathbf{\Phi}_i\mathbf{s}\geq 0\}\times P(\mathbf{\Phi}_i\mathbf{s}\geq 0)\\&\quad+E\{|\mathbf{\Phi}_i\mathbf{s}+\alpha_i|^2\cdot(1-P_i^-)+|\mathbf{\Phi}_i\mathbf{s}-\alpha_i|^2\cdot P_i^-\big|\mathbf{\Phi}_i\mathbf{s}<0\}\times P(\mathbf{\Phi}_i\mathbf{s}<0).\end{aligned}\tag{A.3}$$

Our task is to determine the two conditional expectation terms in (A.3). To obtain the first term, we have

$$=E\{|\mathbf{\Phi}_i\mathbf{s}-\alpha_i|^2\cdot(1-P_i^+)+|\mathbf{\Phi}_i\mathbf{s}+\alpha_i|^2\cdot P_i^+\big|\mathbf{\Phi}_i\mathbf{s}\geq 0\}=E\{(\mathbf{\Phi}_i\mathbf{s})^2-2(1-2P_i^+)\alpha_i\mathbf{\Phi}_i\mathbf{s}+\alpha_i^2\big|\mathbf{\Phi}_i\mathbf{s}\geq 0\}$$

$$\overset{(a)}{=}\int_0^\infty\left[t_i^2-2(1-2P_i^+)t_i\times\alpha_i+\alpha_i^2\right]\times\frac{2\exp(-t_i^2/(2\sigma^2))}{\sqrt{2\pi\sigma^2}}dt_i$$

$$=\alpha_i^2+\int_0^\infty\frac{2t_i^2\exp(-t_i^2/(2\sigma^2))}{\sqrt{2\pi\sigma^2}}dt_i-2\alpha_i\int_0^\infty(1-2P_i^+)t_i\times\frac{2\exp(-t_i^2/(2\sigma^2))}{\sqrt{2\pi\sigma^2}}dt_i$$

$$\overset{(b)}{=}\alpha_i^2+\int_0^\infty\frac{2t_i^2\exp(-t_i^2/(2\sigma^2))}{\sqrt{2\pi\sigma^2}}dt_i-2\alpha_i\int_0^\infty\left[1-2\left(P_{e,i}+(1-2P_{e,i})\times Q\left(t_i/\sigma_v\right)\right)\right]t_i\times\frac{2\exp(-t_i^2/(2\sigma^2))}{\sqrt{2\pi\sigma^2}}dt_i$$

$$\overset{(c)}{=}\alpha_i^2+\sigma^2-2\alpha_i\int_0^\infty\left[1-2\left(P_{e,i}+(1-2P_{e,i})\times Q\left(t_i/\sigma_v\right)\right)\right]t_i\times\frac{2\exp(-t_i^2/(2\sigma^2))}{\sqrt{2\pi\sigma^2}}dt_i$$

$$=\alpha_i^2+\sigma^2-2\alpha_i(1-2P_{e,i})\int_0^\infty\left[1-2Q\left(t_i/\sigma_v\right)\right]t_i\times\frac{2\exp(-t_i^2/(2\sigma^2))}{\sqrt{2\pi\sigma^2}}dt_i,\tag{A.4}$$

where (a) holds since $\mathbf{\Phi}_i\mathbf{s}\sim\mathcal{N}(0,\sigma^2)$ and the density function of $\mathbf{\Phi}_i\mathbf{s}$ conditioned on $\mathbf{\Phi}_i\mathbf{s}\geq 0$ is given by $f_{\mathbf{\Phi}_i\mathbf{s}}(t_i|\mathbf{\Phi}_i\mathbf{s}\geq 0)=\frac{2e^{(-t_i^2/(2\sigma^2))}}{\sqrt{2\pi\sigma^2}}$, (b) is obtained by invoking the definition of $P_i^+$ in (3.8), and (c) follows from[2]

2. The integral in (A.5) yields exactly the variance of a Gaussian random variable with distribution $\mathcal{N}(0,\sigma^2)$.

$$\int_0^{\infty} \frac{2t_i^2 \exp(-t_i^2/(2\sigma^2))}{\sqrt{2\pi\sigma^2}} dt_i = \int_{-\infty}^{\infty} \frac{t_i^2 \exp(-t_i^2/(2\sigma^2))}{\sqrt{2\pi\sigma^2}} dt_i = \sigma^2. \quad \text{(A.5)}$$

An explicit form of the remaining integral term in (A.4) can be obtained as

$$\begin{aligned}
&\int_0^{\infty} [1 - 2Q(t_i/\sigma_v)] t_i \times \frac{2\exp(-a_i^2/(2\sigma^2))}{\sqrt{2\pi\sigma^2}} dt_i \\
&= \int_0^{\infty} t_i \times \frac{2\exp(-t_i^2/(2\sigma^2))}{\sqrt{2\pi\sigma^2}} dt_i - 2\int_0^{\infty} t_i Q(t_i/\sigma_v) \times \frac{2\exp(-t_i^2/(2\sigma^2))}{\sqrt{2\pi\sigma^2}} dt_i \\
&= \sqrt{2\sigma^2/\pi} - 2\int_0^{\infty} t_i Q(t_i/\sigma_v) \times \frac{2\exp(-t_i^2/(2\sigma^2))}{\sqrt{2\pi\sigma^2}} dt_i \\
&\overset{(a)}{=} \sqrt{2\sigma^2/\pi} - 2\sqrt{\sigma^2/(2\pi)}\left(1 - \sqrt{\sigma^2/(\sigma^2 + \sigma_v^2)}\right) = \sqrt{2\sigma^2/\pi} \times \sqrt{\sigma^2/(\sigma^2 + \sigma_v^2)},
\end{aligned} \quad \text{(A.6)}$$

where (a) follows from Lemma A.1. With (A.4) and (A.6), it follows that

$$\begin{aligned}
&E\{|\mathbf{\Phi}_i \mathbf{s} - \alpha_i|^2 \cdot (1 - P_i^+) + |\mathbf{\Phi}_i \mathbf{s} + \alpha_i|^2 \cdot P_i^+ \,|\, \mathbf{\Phi}_i \mathbf{s} \geq 0\} \\
&= \alpha_i^2 + \sigma^2 - 2\alpha_i(1 - 2P_{e,i}) \times \sqrt{2\sigma^2/\pi} \times \sqrt{\sigma^2/(\sigma^2 + \sigma_v^2)} \\
&= \alpha_i^2 - 2\alpha_i(1 - 2P_{e,i}) \times \sqrt{2\sigma^2/\pi} \times \sqrt{\sigma^2/(\sigma^2 + \sigma_v^2)} + \sigma^2.
\end{aligned} \quad \text{(A.7)}$$

By going through essentially the same procedures, it can be shown that

$$\begin{aligned}
&E\{|\mathbf{\Phi}_i \mathbf{s} + \alpha_i|^2 \cdot (1 - P_i^-) + |\mathbf{\Phi}_i \mathbf{s} - \alpha_i|^2 \cdot P_i^- \,|\, \mathbf{\Phi}_i \mathbf{s} < 0\} \\
&= \alpha_i^2 - 2\alpha_i(1 - 2P_{e,i}) \times \sqrt{2\sigma^2/\pi} \times \sqrt{\sigma^2/(\sigma^2 + \sigma_v^2)} + \sigma^2.
\end{aligned} \quad \text{(A.8)}$$

Using (A.3), (A.7), (A.8), and since $\Pr\{\mathbf{\Phi}_i \mathbf{s} \geq 0\} = \Pr\{\mathbf{\Phi}_i \mathbf{s} < 0\} = 1/2$, we have

$$E\{|w_i|^2\} = \left(\alpha_i - \sqrt{\frac{2\sigma^2(1 - 2P_{e,i})^2}{\pi \cdot (1 + (\sigma_v/\sigma)^2)}}\right)^2 + \sigma^2 - \frac{2\sigma^2(1 - 2P_{e,i})^2}{\pi(1 + (\sigma_v/\sigma)^2)}, \quad \text{(A.9)}$$

and (3.9) follows immediately. □

# REFERENCES

[1] R. G. Baraniuk, "Compressive sensing," *IEEE Signal Processing Magazine*, vol. 24, no. 4, pp, 118-124, July 2007.

[2] Y. C. Eldar and G. Kutyniok, ed., *Compressed Sensing: Theory and Applications*, Cambridge University Press, 2011.

[3] J. N. Laska, P .T. Boufounos, M. A. Davenport, and R. G. Baraniuk, "Democracy in action: Quantization, saturation, and compressive sensing," *Applied Computational and Harmonic Analysis*, vol. 31, pp. 429-443, 2011.

[4] P. T. Boufounos and R. G. Baraniuk, "1-bit compressive sensing," *Proc. 42nd Annual Conf. on Information Sciences and Systems*, Princeton, NJ, March 2008, pp. 16-21.

[5] L. Jacques, J. N. Laska, P. T. Boufounos and R. G. Baraniuk, "Robust 1-bit compressive sensing via binary stable embedding of sparse vectors," *IEEE Trans. Information Theory*, vol. 59, no. 4, pp. 2082-2202, April 2013.

[6] M. Yan, Y. Yang, and s. Osher, "Robust 1-bit compressive sensing using adaptive outlier pursuit," *IEEE Trans. Signal Processing*, vol. 60, no. 7, pp. 3868-3875, July 2012.

[7] A. Movahed, A. Panahi, and G. Durisi, "A robust RFPI-based 1-bit compressive sensing recovery algorithm," *Proc. IEEE Information Theory Workshop*, 2012, pp. 567-571.

[8] J. D. Haupt, W. U. Bajwa, M. Rabbat, and R. D. Nowak, "Compressed sensing for networked data," *IEEE Signal Processing Magazines*, vol. 25, no. 2, pp. 92-101, March 2008.

[9] A. Y. Yang, M. Gastpar, R. Bajcsy, and S. S. Sastray, "Distributed sensor perception via sparse representation," *Proc. of the IEEE*, vol. 98, no., 6, pp. 1077-1088, June 2010.

[10] T. Sakdejayont, D. Lee, Y. Peng, Y. Yamashita, and H. Morikawa, "Evaluation of memory-efficient 1-bit compressed sensing in wireless sensor networks, *Proc. 2013 IEEE Region 10 Humanitarian Technology Conference*, Sendai, Japan, Aug. 2013, pp. 326-329.
[11] Y. Shen, J. Fang, and H. Li, "One-bit compressive sensing and source location is wireless sensor networks" *Proc. 2013 China Summit & International Conference on Signal and Information Processing*, pp. 379-383.
[12] E. Candes, "The restricted isometry property and its implications for compressed sensing," *C. R. Acad. Sci. Paris, Ser I* 346, pp. 589-592, 2008.
[13] Z. Yan, K. M. Wong, and Z. Q. Luo, "Optimal diagonal precoder for multiantenna communication systems," *IEEE Trans. Signal Processing*, vol. 53, no. 6, pp. 2089-2100, June 2005.